# Bipolaronic nature of the pseudogap in (TaSe$_4$)$_2$I revealed via weak photoexcitation

Yingchao Zhang[1*], Tika Kafle[1], Wenjing You[1], Xun Shi[1], Lujin Min[3], Huaiyu(Hugo) Wang[3], Na Li[1], Venkatraman Gopalan[3], Kai Rossnagel[7,8], Lexian Yang[5,6], Zhiqiang Mao[3,4], Rahul Nandkishore[2], Henry Kapteyn[1], Margaret Murnane[1]

[1]*Department of Physics and JILA, University of Colorado and NIST, Boulder, CO 80309, USA*
[2]*Department of Physics and Center for Theory of Quantum Matter, University of Colorado Boulder, Boulder, CO 80309, USA*
[3]*Materials Research Institute and Department of Materials Science & Engineering, The Pennsylvania State University, University Park, PA 16802, USA*
[4]*Department of Physics, The Pennsylvania State University, University Park, PA 16802, USA*
[5]*State Key Laboratory of Low Dimensional Quantum Physics, Department of Physics, Tsinghua University, Beijing 100084, China.*
[6]*Frontier Science Center for Quantum Information, Beijing 100084, China.*
[7]*Institute of Experimental and Applied Physics, Kiel University, D-24098 Kiel, Germany*
[8]*Ruprecht Haensel Laboratory, Deutsches Elektronen-Synchrotron DESY, D-22607 Hamburg, Germany*

\* Corresponding author: Yingchao.Zhang-1@colorado.edu

The origin of the pseudogap in many strongly correlated materials has been a longstanding puzzle. Here, we uncover which many-body interactions underlie the pseudogap in quasi-one-dimensional (quasi-1D) material (TaSe$_4$)$_2$I by weak photo-excitation of the material to partially melt the ground state order and thereby reveal the underlying states in the gap. We observe the appearance of both dispersive and flat bands by using time- and angle-resolved photoemission spectroscopy. We assign the dispersive band to a single-particle bare band, while the flat band to a collection of single-polaron sub-bands. Our results provide direct experimental evidence that many-body interactions among small Holstein polarons i.e., the formation of bipolarons, are primarily responsible for the pseudogap in (TaSe$_4$)$_2$I. Recent theoretical studies of the Holstein model support the presence of such a bipolaron-to-polaron crossover. We also observe dramatically different relaxation times for the excited in-gap states in (TaSe$_4$)$_2$I (~600 fs) compared with another quasi-1D material Rb$_{0.3}$MoO$_3$ (~60 fs), which provides a new method for distinguishing between pseudogaps induced by polaronic or Luttinger-liquid many-body interactions.

**Introduction**

The absence of a clear Fermi edge along with the existence of a pseudogap in a broad class of materials including cuprates, colossal magnetoresistance manganites and quasi-one-dimensional (quasi-1D) materials has been a long standing and important puzzle. For example, it is believed to play a vital role in a series of strongly correlated phenomena including high-temperature superconductivity(*1–4*). Strong correlations in combination with multiple mechanisms such as polaronic interactions(*1, 3*), Luttinger-liquid behavior(*4, 5*), Efros-Shklovski effect(*6, 7*) and charge density wave (CDW) fluctuations(*8*) can be responsible for opening a pseudogap. Determining the dominant mechanism leading to the formation of a pseudogap is challenging, partly due to the lack of characteristic features to distinguish between different mechanisms. This is the case in many conventional static equilibrium measurements such as angle-resolved photoemission spectroscopy (ARPES), electrical transport and optical conductivity(*3, 5, 9, 10*). These techniques share the characteristic of probing only the electronic bands close to the ground state. For instance, many potential mechanisms are expected to give rise to fine features in the ARPES spectral function in the pseudogap i.e. a quasi-particle peak and polaron sub-bands(*11, 12*). However, these features are often difficult to observe under equilibrium conditions because of the strongly suppressed spectral weight within the pseudogap. Past work demonstrated that the timescale for melting the order can yield useful information for distinguishing the dominant interactions in strongly-coupled materials(*13*). Here we use weak laser excitation to partially melt the ground state order and reveal the underlying states in the pseudogap, essentially driving a transient crossover between two different many-body regimes. Using time-resolved ARPES, we measure both the energy-momentum distribution and characteristic formation and relaxation times of these *emergent excited in-gap states*, that provide insight into the dominant interactions that underlie *the ground state*. This approach provides clear signatures of the nature of the many-body interactions leading to a pseudogap, making it possible to distinguish pseudogaps induced by the formation of bipolarons(*9*), or the formation of a Luttinger liquid(*5*).

Quasi-1D materials are excellent platforms for observing strongly correlated phenomena such as CDW order(*14–16*) and Luttinger liquid behavior(*4, 5*) since they exhibit a reduced phase space for scattering and less screening. Recently, quasi-1D (TaSe$_4$)$_2$I has attracted much attention since it may become an axion insulator when the Weyl points are modified by CDW

order(*17, 18*). However, to date no clear evidence of Weyl points has been observed, partly due to the missing spectral weight at the Fermi surface of $(TaSe_4)_2I$. Compared to 3D materials, the Fermi surface in quasi-1D materials is more susceptible to Peierls interactions, which open a CDW gap at the Fermi level. However, even in the normal phase above the CDW transition temperature, a clear Fermi edge is still absent in most quasi-1D materials – including $(TaSe_4)_2I$. In some quasi-1D materials such as $Li_{0.9}Mo_6O_{17}$ and $K_{0.3}MoO_3$(*4, 5*), the opening of a pseudogap in the normal phase could be attributed to the formation of a Luttinger liquid. CDW and magnetic fluctuations could also reduce the spectral weight at the Fermi surface, especially when the scattering phase-space is reduced in 1D (*8, 9*). However, these mechanisms alone cannot explain the ARPES spectra of $(TaSe_4)_2I$ observed above and below the CDW phase transition.

Polaron formation has been proposed as a mechanism for the pseudogap opening in colossal magnetoresistance manganites(*1*). A polaron consists of a carrier dressed by a cloud of virtual phonons(*19, 20*). In $(TaSe_4)_2I$(*3*), the strong electron-phonon coupling and large effective mass of the carriers suggest that small Holstein polarons could play an important role in determining the low energy excitations. For decades, many efforts have been made to solve the Holstein model with many-body interactions in different dimensions. Recent Monte Carlo simulations of the Holstein model at the adiabatic limit have produced a complex phase diagram with CDW and bipolaron insulating phases, as well as a metallic single-polaron phase, as shown schematically in Fig. 1A(*21*).

Here we report clear evidence of a photoinduced bipolaron-to-polaron crossover in $(TaSe_4)_2I$ by using trARPES to probe the electronic structure of a gently photo-doped excited state. After excitation by a femtosecond laser, two new bands are revealed in the pseudogap of $(TaSe_4)_2I$ - a straight-line dispersive band and a non-dispersive flat band. The new dispersive band resembles the single-particle bare band dispersion (dashed black line in Fig. 1B). The transiently-revealed flat band observed in the trARPES data, which does not appear in the calculated single-particle band structure(*18, 22, 23*), is consistent with a collection of smeared single-polaron sub-bands within the pseudogap, as illustrated schematically in Fig. 1B (solid black line). The emergence of these two bands after partial melting of the bipolaron states provides direct evidence that many-body polaronic interaction - *namely bipolaron formation* - reduces the spectral weight of the single-particle bare band at the Fermi surface and opens up a polaron pseudogap. The relatively weak laser excitation ensures that we only partially melt the bipolaron insulator into a single-polaron metal, and avoid driving the material into a normal metal state.

To extract more information about the many-body interactions, we also track the ultrafast evolution of the spectral weight in the pseudogap. The dynamics of the in-gap states show that the bipolarons are partially melted into single polarons within ~250 fs, before gradually recovering with a time constant of ~600 fs – which is a typical relaxation time for photoinduced structural changes(*13*, *24*). These findings, when combined with recent theoretical predictions(*21*), provide compelling evidence that polaronic many-body interactions play a key role in opening the pseudogap in (TaSe$_4$)$_2$I. Moreover, we conduct the trARPES measurements for another prototypical quasi-1D material Rb$_{0.3}$MoO$_3$ that exhibits dramatically faster (~10x) relaxation. The pseudogap in this material was variously attributed either to the presence of a polaron gap(*9*), or the formation of a Luttinger liquid(*5*). Surprisingly, we find that the decay constant of the excited in-gap states in Rb$_{0.3}$MoO$_3$ is only ~60 fs. This fast relaxation is likely electronic in nature – such as hot carriers losing energy to Luttinger plasmons. Thus, our results favor the recent Luttinger-liquid explanation of the pseudogap in Rb$_{0.3}$MoO$_3$(*5*). Our data indicate that the polaronic effects previously proposed(*9*) are *not* crucial to the pseudogap physics in Rb$_{0.3}$MoO$_3$. This work thus represents a new approach for uncovering the dominant many-body interactions via weak ultrafast photo-doping, which may help explain a set of puzzling strongly correlated phenomena such as the strange metal phase in high-temperature superconductors.

1. Results

(TaSe$_4$)$_2$I is a prototypical quasi-1D CDW material which enters an incommensurate CDW phase at temperatures below 263 K ($T_{CDW}$). Theoretical calculations propose that it is a Weyl semimetal above $T_{CDW}$ and that it becomes an axion insulator below $T_{CDW}$(*18*, *23*). (TaSe$_4$)$_2$I has a conventional tetragonal unit cell ($a = b = 9.531$Å and $c = 12.824$ Å). The Ta atoms form chains, surrounded by Se$_4$ rectangular units and separated by iodine ions. Two adjacent Se$_4$ rectangles are rotated by 45°, which makes TaSe$_4$ chains exhibit a screwlike symmetry – thus, (TaSe$_4$)$_2$I is known as a chiral crystal(*22*). The chains are bonded weakly by iodine atoms, forming needle-like crystals that naturally cleave along the (110) plane. The second material we study is Rb$_{0.3}$MoO$_3$, also known as blue bronze, which is another prototypical quasi-1D CDW material ($T_{CDW}\sim$183 K) that has similar properties to (TaSe$_4$)$_2$I(*3, 9, 25, 26*).

We perform trARPES measurements on (TaSe$_4$)$_2$I at both room (300 K) and low temperatures (80 K). For our measurements, the chain direction is aligned along the analyzer slit. We excite the materials using a 1.6 eV infrared laser pump pulse (~40 fs, 10 kHz), and probe the dynamic electronic order using a 22 eV extreme ultraviolet (EUV) high harmonic probe pulse (~10 fs, 10 kHz). The time and energy resolution are <10 fs and ~100 meV, respectively. Figure 2 shows the raw trARPES spectra at different temperatures and time delays. The in-plane band dispersion (Fig. 2A) shows a characteristic "V"-shaped valence band (VB). The VB maximum is at around $1.1\pi/c$, with ~0.4 eV binding energy. The conduction band (CB) minimum is below the Fermi level due to natural n-type doping from iodide vacancies. A CDW gap ($2\Delta_{CDW}\approx$0.2 eV) is opened between the VB maxima and CB minima below 263 K(*27*). In addition to the CDW gap, a pseudogap ($2\Delta_{BP}\approx$0.4 eV) removes most of the spectral weight of the CB at the Fermi level. Most importantly, this pseudogap is present in both the low temperature CDW phase and the room temperature normal phase, as shown in Figs. 2A and B.

Figures 3A-E plot the second derivative images of the raw data at different temperatures and time delays in Figure 2. The second derivative along the vertical direction is best for enhancing the horizontally dispersing band, while the horizontal second derivative is best for enhancing the vertically dispersing band. A straight-line-like dispersive band appears in the pseudogap approximately 250 fs after laser excitation, as shown in Figs. 3B, C and E. The room temperature and low temperature data show very similar behaviors for the dynamics of this new band. The dispersion of the new band is tracked by fitting the momentum distribution curves (MDC), which are presented in Fig. 3H. Due to the transition matrix effect, we can only resolve one branch of the Dirac cone. Thus, a single Lorentzian peak is applicable to fit the band dispersion. This new band can be recognized as the single-particle bare band – which is not present in the data before laser excitation due to polaronic interactions. The band velocity, which is almost constant for all positive time delays (Fig. S3A), is determined to be 7.5 eVÅ – almost twice as the VB velocity between 0.1 Å$^{-1}$ and 0.2 Å$^{-1}$. Note that the room temperature spectra before laser excitation presented in Figs. 3A and H still show a weak Fermi crossing with a very high band velocity (~30 eVÅ). The band velocities derived from MDC fittings can be influenced by artifacts

resulting from the way MDC analysis handles remnants of the broad hump in the pseudogap(*11*). We believe the large unpumped band velocity (~30 eVÅ) is precisely such an artifact. However, we believe the smaller 7.5 eVÅ velocity observed after pump excitation is meaningful(*11*), and marks the partial 'undressing' of polarons into bare electrons. This critical change of the band velocity can be directly seen from the MDC curves of the raw data, presented in Figs. 3F and G.

In addition to revealing the dispersive bare band, another non-dispersive flat band also emerges near the Fermi level (–0.16 eV) after pump excitation, as denoted by black dashed lines in Figs. 4A-D. Since this weak band is on the shoulder of an intense broad hump, it is not clearly recognized from the raw and second derivative images in Figs. 2 and 3. Nevertheless, the difference maps (Figs. 4A at 80 K and 4C at 300 K) between 250 fs and the negative time delay can remove the intense background and enhance photo-induced weak features, which helps us to recognize the photo-enhanced flat band centered at –0.16 eV. Figures 4B and D show the second derivative of the images in Figs. 4A and C respectively, in which the flat band is enhanced throughout the entire Brillouin zone. More error analysis can be found in the supplementary material regarding overcoming artifacts posed by band broadening and shifting. The net gain of the spectral weight at 0.5 Å between -1 eV and 0.5 eV, as shown in Fig. S2B, is strong evidence that the broadening and up-shift of the broad valence band at higher binding energy are not the major origins of the flat band centered at –0.16 eV, as both of them conserve the spectral weight. The flat band can be explained as a collection of single-polaron sub-bands within the pseudogap, which is consistent with the calculated spectral function from single-electron Holstein mode(*28*).

Furthermore, we track the dynamics of the spectral weight within the pseudogap of both $(TaSe_4)_2I$ and $Rb_{0.3}MoO_3$, as denoted by the yellow dashed rectangles in Figs. 4C and S6. Figure 4E plots the excited state dynamics of $(TaSe_4)_2I$ and $Rb_{0.3}MoO_3$ for a pump fluence of 0.6 mJ/cm$^2$ at 300 K. For $(TaSe_4)_2I$, upon excitation, the electrons in the VB are pumped into the CB and are thermalized around the Fermi level within ~100 fs, with the electron temperature rising to a peak value of ~1600 K (Fig. S3B). The spectral weights within the pseudogap start to appear immediately after the arrival of the pump pulse and reach their peak value at ~250 fs. Then the ARPES spectra gradually recover to the ground state within ~1 ps. The dynamics of the in-gap states for the CDW phase at 80 K and the normal phase at 300 K are similar (Fig. S6), which indicates that polaron dynamics dominate CDW dynamics at this 0.6 mJ/cm$^2$ pump fluence. We believe that the rapidly excited hot carriers screen the bipolaronic interactions within 50 fs, while it takes ~ 250 fs for the two interacting polarons in a bipolaron to fall apart, due to the bottleneck imposed by the speed of the lattice relaxation. The screened bipolaronic interactions gradually recover following heat transfer from the hot electron bath to the cold phonon bath. The photo-generated single polarons remerge into bipolarons with a time constant of ~600 fs.

In the case of $Rb_{0.3}MoO_3$, after excitation by a 1.6 eV photon-energy pump pulse, electron-hole pairs are excited in the region above the Fermi level. Then these hot carriers quickly lose their energy via emitting phonons and relax to the vicinity of the Fermi level within 100 fs, where Luttinger-liquid interactions start to dominate. These excited states relax with a 10x faster time constant of ~60 fs, possibly by emitting Luttinger plasmons. We note that similarly fast time scales have been observed in previous trARPES measurements of $Rb_{0.3}MoO_3$(*29*).

## 2. Discussion

With the new experimental findings presented in this paper, we can now clarify the long-standing challenges of explaining the opening of a pseudogap in quasi-1D $(TaSe_4)_2I$. This methodology can also be extended to solve the puzzle of missing Fermi surface states in other quasi-1D materials. Here we discuss four possible mechanisms: (i) Luttinger-liquid behavior; (ii) the Efros-Shklovski effect; (iii) fluctuations of the CDW order parameter; (iv) bipolaronic interactions. With strong electronic correlations in a 1D metal, single-particle excitations can be rearranged into collective bosonic excitations, which form a Luttinger liquid with various peculiar features - including power-law singularity and spin-charge separation. The power-law singularity of the spectral function in the vicinity of the Fermi level induces a pseudogap such that single-particle excitations are greatly suppressed. Nevertheless, the power-law distribution of the spectral function does not modify the band dispersion(*5*). Thus, the change in band velocity after pump excitation in $(TaSe_4)_2I$ (Figs. 3F-H) is clearly not a signature of Luttinger liquid behavior.

The strong electron-phonon coupling that enhances the effective mass of the carriers can make them more susceptible to Anderson localization. The static Coulomb repulsions among Anderson-localized carriers render the so-called Efros-Shklovski pseudogap(*6, 7*). However, there is not a natural way for a disorder-localized high temperature phase to give rise to a low temperature CDW phase. The well-defined CDW phase with characteristic phason-induced nonlinear resistivity and angular dependence of the axial current in (TaSe$_4$)$_2$I at low temperature does not support a dominant role of Anderson localization in the electronic properties(*17*). With reduced scattering phase-space, fluctuations can be more dominant for lower dimensionality. Random CDW regions above $T_{CDW}$ could reduce the spectral weight on the Fermi surface and open up a pseudogap(*8, 9*). However, the CDW gap is below the Fermi level in the naturally doped crystal and smaller than the pseudogap(*27*), which rules out the possibility that the pseudogap is mainly induced by CDW fluctuations.

Previous ARPES and optical conductivity measurements have suggested a possible polaronic origin of the pseudogap in (TaSe$_4$)$_2$I(*3*). However, direct experimental observations of the influence of polaronic effects are still needed. Although simulating many-electron Holstein polarons is still challenging, the one-electron Holstein model, with a single electron coupled to an Einstein phonon mode (with $\hbar\omega_0$ phonon energy), has been solved and results in a spectral function that exhibits kinks, and multiple polaron sub-bands at $n\hbar\omega_0$ binding energy ($n = \pm 1, \pm 2 ...$)(*28*). Although the calculated spectral functions of one-electron Holstein polarons agree well with a series of ARPES measurements(*5, 20, 30*), in many materials with strong electron-phonon coupling including (TaSe$_4$)$_2$I, the large pseudogap (~200 meV) cannot be reproduced by a one-electron Holstein model. Instead, many-body interactions among single polarons need to be taken into account in order to generate such a pseudogap. Previous calculations using a many-electron Holstein model in the atomic limit ($\hbar\omega_0 \gg t$, where $t$ is the electron hopping integral) can well reproduce the pseudogap features, bearing similarity to the Franck-Condon lineshape(*12, 31*). However, in (TaSe$_4$)$_2$I, the adiabatic limit ($\hbar\omega_0 \ll t$, with $\hbar\omega_0 < 35$ meV, $t \sim 1$ eV) is more applicable. Recent Monte Carlo simulations of the Holstein model in the adiabatic limit show a complex phase diagram in the $\lambda - T$ plane (where $\lambda$ is the dimensionless electron-phonon coupling strength and $T$ is the temperature), as illustrated schematically in Fig. 1A(*21*). Above the CDW region, there exists not only a metallic regime but also an insulating regime, which is associated with a gap or pseudogap across the Fermi level. A clear phase crossover between the two regimes is controlled by $\lambda$ and $T$. A simple microscopic description of the insulating regime considers two metallic single polarons that can merge into a bipolaron and gain energy $U = g^2/\omega_0^2 = \lambda/N$, where $g$ is the electron-phonon coupling constant and $N$ is the bare density of state at the Fermi level. The lattice can be seen as a binary alloy with $-U$ bipolaron sites and $+U$ remaining sites. A gap of $2U$ is thus opened due to the formation of bipolarons. And in (TaSe$_4$)$_2$I, a few phonon modes have sufficiently large electron-phonon coupling strength $\lambda$ to generate bipolarons with a few hundred meV binding energy(*10*). However, we do not have direct evidence to determine which phonon mode is responsible for bipolaron formation. Nevertheless, empirically, the bipolaron melting time (~250 fs) should correspond to ~1/2 period of the relevant phonon mode(s), which suggests a phonon frequency of ~2 THz. Our time-resolved optical reflectivity measurements identify two candidate phonon modes with frequencies close to this (2 THz and 2.5 THz) and strong responses to optical excitation (Fig. S5). A previous study(*10*) also reports phonon modes with similar frequencies and large electron-phonon coupling strength $\lambda$ that could produce a ~200 meV bipolaron pseudogap. Nevertheless, more conclusive studies are needed to finally determine the phonon mode that generates the bipolaron pseudogap.

This bipolaron picture can explain the pseudogap observed in the ARPES spectra as well as the activation behavior of conductivity in (TaSe$_4$)$_2$I at $T > T_{CDW}$. After laser excitation, the electron temperature increases to 1600 K within ~50 fs. The resulting hot electron distribution, although low density, can reduce the electron-phonon coupling strength $\lambda$ and screen the bipolaronic interactions. The relatively high plasma frequency (~126 THz at 300 K(*3*)) ensures that the effective screening can build up within 10 fs. Within ~250 fs, the bipolarons are broken into free single polarons, which results in a large number of single-polaron states in the vicinity of the Fermi level. We note that the ~250 fs rise time of the excited single-polaron states remains unchanged when tuning the pump fluence from 0.1 to 0.6 mJ/cm$^2$ (Fig. S8), which indicates that this characteristic rise time results from a structural bottleneck imposed by the half-period of the strongly coupled phonon mode. The single-polaron states are derived from the bare band hybridizing with the strongly coupled phonon mode, which forms a series of discrete polaron sub-bands, as shown schematically in Fig. 1B (solid black line). Each polaron sub-band has two non-dispersive branches extending over the whole Brillouin zone at the two sides of the original bare band. The spacing between two adjacent polaron sub-bands is equal to $\hbar\omega_0$ which should be less than 35 meV (the maximum phonon energy in (TaSe$_4$)$_2$I). Thus, we cannot resolve an individual polaron sub-band with ~100 meV energy resolution. Instead, we observe one flat band throughout the entire Brillouin zone within the gap, which corresponds to a collection of several smeared non-dispersive branches of the polaron sub-bands. The excess energy stored in the electron bath and strongly coupled phonon mode can then be transferred and thermalized to the rest of the phonon bath in ~1 ps, when the single polarons relax into bipolarons.

The relaxation timescales also yield useful information for distinguishing different many-body interactions that underlie the pseudogap. For example, the identical experiment in another 1D material $Rb_{0.3}MoO_3$ yields a much faster relaxation timescale of ~60 fs (Fig. 4E). The order of magnitude difference in relaxation times (60 fs vs 600 fs) suggests a different underlying mechanism, which in turn suggests that bipolaron effects are likely *not* crucial to the pseudogap physics in $Rb_{0.3}MoO_3$. We speculate that Luttinger liquid effects may be responsible instead, with relaxation occurring through emission of Luttinger plasmons. When the initial electron-hole pairs excited by 1.6 eV photons thermalize to the vicinity of the Fermi level, they can rearrange into collective bosonic excitations (Luttinger plasmons) that are nearly invisible in ARPES measurements. This decay channel could be much faster than the bipolaron reformation, since it is not bottlenecked by the lattice structural relaxation. We note that a recent high-resolution static equilibrium ARPES study(*5*) has also observed power law singularities consistent with Luttinger liquid physics in this material. Other mechanisms, such as heat transfer from hot carriers to strongly coupled high-energy optical phonon modes(*29*, *32*), could also contribute to the fast relaxation of excited states in $Rb_{0.3}MoO_3$. Nevertheless, the ~60 fs relaxation time can still help to confirm the existence of dominant Luttinger Liquid physics, since it rules out pseudogap mechanisms involving structural phase changes such as bipolaron effects and CDW fluctuations.

In summary, we provide direct evidence that bipolaron formation in $(TaSe_4)_2I$ plays a major role in the formation of the pseudogap. This is supported by recent Monte Carlo simulations of the Holstein model. The ultrafast in-gap dynamics observed depict a picture in which bipolarons are transiently broken into single polaron states by screening by hot electrons, forming again once the electron subsystem cools. By comparing the ultrafast dynamics of $(TaSe_4)_2I$ and $Rb_{0.3}MoO_3$, we show that our new approach for determining the dominant many-body interactions can be extended to understand the nature of the pseudogap phase in other materials. In the future, this approach can help with understanding a set of puzzling strongly correlated phenomena such as the strange metal phase.

## ACKNOWLEDGEMENTS

We thank Steve Kivelson for feedback on the manuscript. RN also thanks Chaitanya Murthy for several stimulating conversations. Y.Z., T.K., W.Y., X.S., H.K. and M.M. gratefully acknowledge support from the National Science Foundation through the JILA Physics Frontiers Center PHY- 1734006. This material is based in part (RN) upon work supported by the Air Force Office of Scientific Research under award number FA9550-20-1-0222. RN also acknowledges the support of the Alfred P. Sloan Foundation through a Sloan Research Fellowship, and of the Simons Foundation through a Simons Fellowship in Theoretical Physics. L.J.M. and Z.Q.M. acknowledge the support from NSF through the Materials Research Science and Engineering Center DMR 2011839. H.W. and V. G. acknowledge the Department of Energy DE-SC0012375 for primary support, and DE-SC0020145 for partial support.

## AUTHOR DECLARATIONS

**Conflict of Interest:** H.K. and M.M. have a minority financial interest in a laser company, KMLabs, that produces engineered versions of the lasers and HHG sources used in this work. H.K. is partially employed by KMLabs. The authors declare that they have no other competing interests.

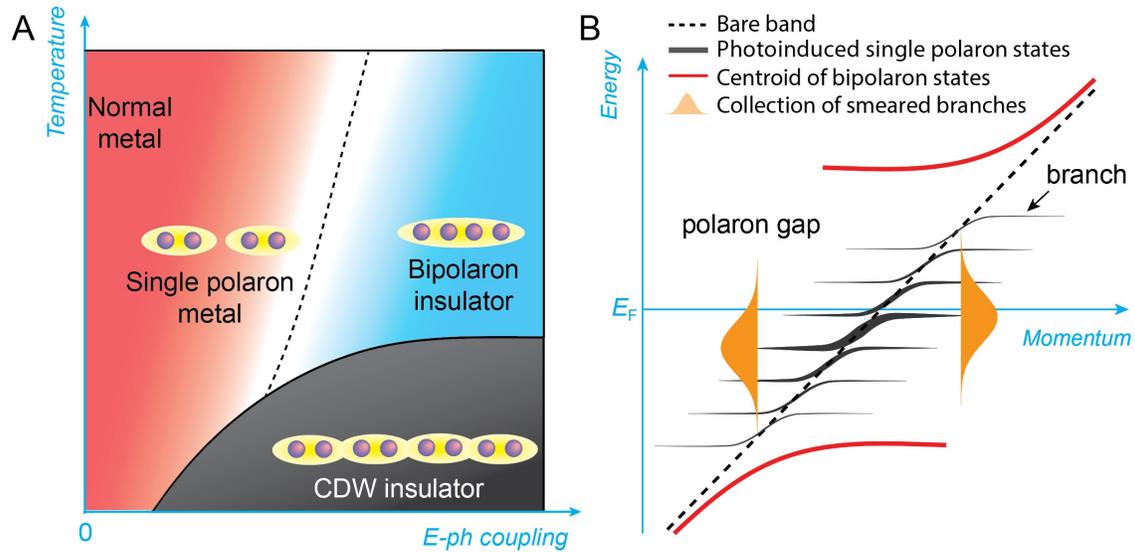

**Fig. 1. (A)** Schematic of the $\lambda - T$ phase diagram based on Monte Carlo simulations of Holstein model at adiabatic limit(*21*). **(B)** Schematic of the predicted static photoemission spectra in the presence of strong electron-phonon coupling and bipolaron formation. The in-gap states are the single-polaron sub-bands formed by bare electronic band coupling to a non-dispersive phonon mode in single-electron Holstein model. Each single-polaron polaron sub-band has two non-dispersive branches extending over the whole Brillouin zone at the sides of the original bare band. The adjacent polaron sub-bands are spaced by the energy of the strongly coupled phonon mode. The flat band we transiently observe within the pseudogap after laser excitation represents a collection of side branches from several polaron sub-bands.

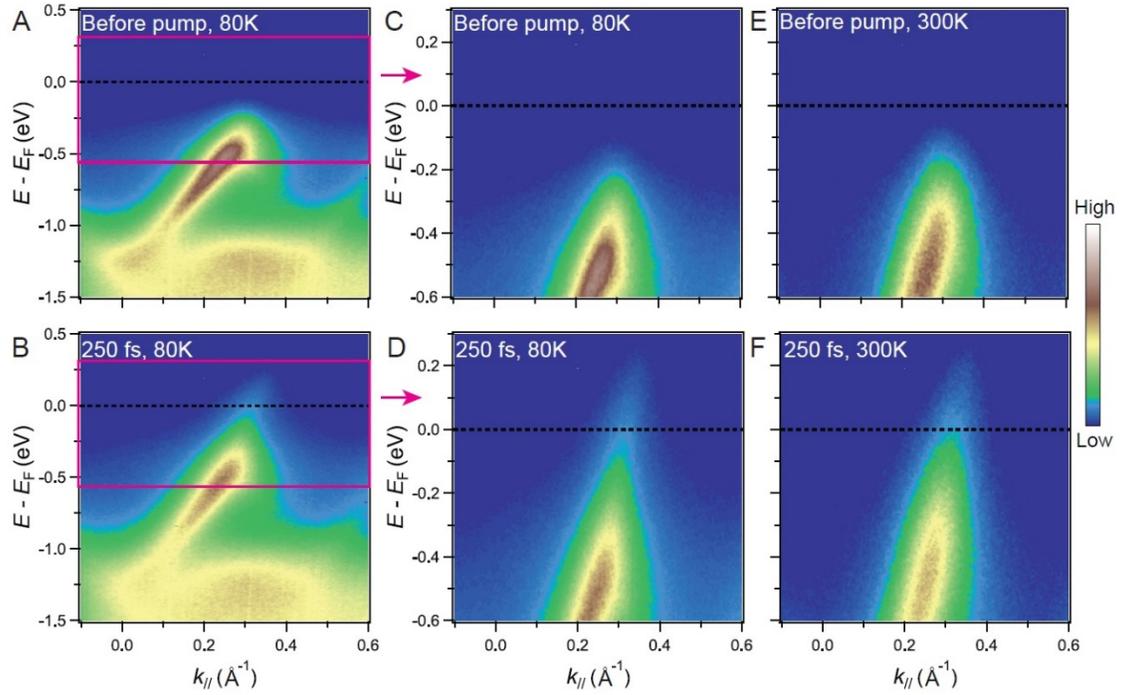

**Fig. 2. (A)** to **(F)** show the ARPES mappings at different temperatures and time delays. **(C)** and **(D)** are the zoom-in of the areas enclosed by the pink rectangles in **(A)** and **(B)**, respectively.

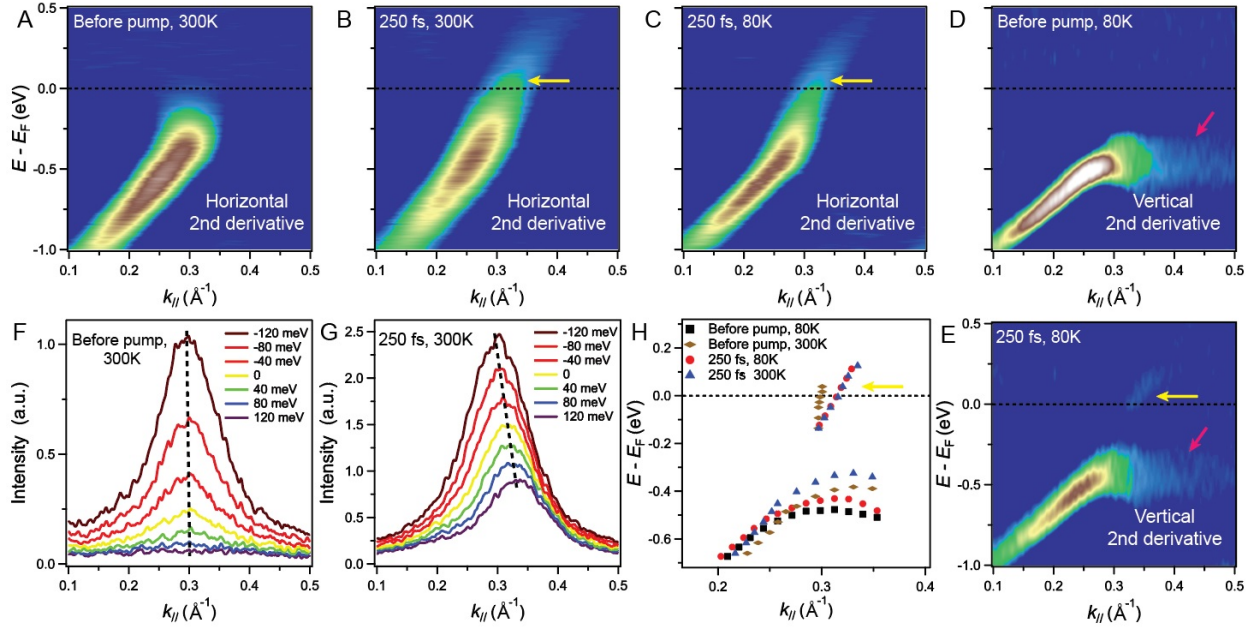

**Fig. 3. (A)** to **(E)** show the vertical and horizontal second derivative images of the corresponding raw ARPES data at different temperatures and time delays, for 0.6 mJ/cm$^2$ pump fluence. The yellow arrows denote the newly emerged dispersive bare band (dashed black line in Fig. 1), while the pink arrows denote the renormalized valence band dispersion at large momenta (lower red band in Fig. 1B). **(F)** and **(G)** present the MDC curves in the vicinity of Fermi level (±120 meV) extracted from the raw ARPES data at 300 K, before and after pump excitation, respectively. **(H)** shows the corresponding centroid of the band dispersion at different time delays and temperatures. Here the MDCs were fit for the vertically trending part, while the EDC curves were fit for the horizontally trending part.

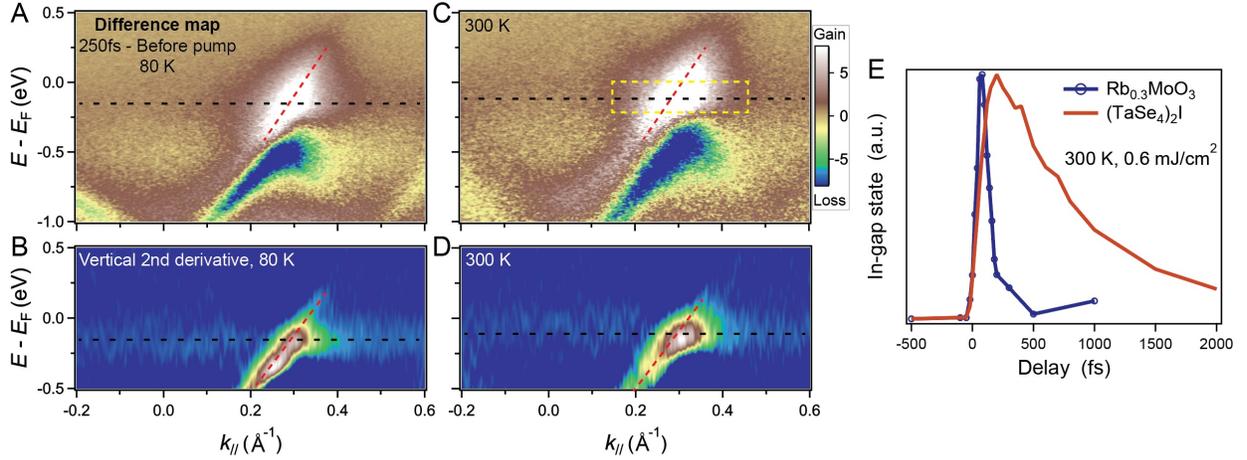

**Fig. 4. Observation of coherent single-polaron flat bands in (TaSe$_4$)$_2$I and dramatically different relaxation times for the excited in-gap states in (TaSe$_4$)$_2$I and Rb$_{0.3}$MoO$_3$.** **(A)** and **(B)** show the difference maps between ARPES spectra at 250 fs and at negative time delay at 80 K and 300 K, respectively. **(B)** and **(D)** are the corresponding vertical second derivative of the difference maps in **(A)** and **(C)**. The red dashed lines track the dispersion of the photo-excited dispersive bare band, while the black dashed lines track the non-dispersive flat bands which can be assigned to a collection of single-polaron sub-bands (solid black bands in Fig. 1). **(E)** show the dynamics of the in-gap spectral weight within a region defined by the yellow dashed rectangles in **(C)** for (TaSe$_4$)$_2$I and in Fig. S7 for Rb$_{0.3}$MoO$_3$. The decay constants of 649 ± 31 fs and 65 ± 8 fs for (TaSe$_4$)$_2$I and Rb$_{0.3}$MoO$_3$ respectively are obtained by single-exponential fittings.

# Supplementary Materials for

# Bipolaronic nature of the pseudogap in $(TaSe_4)_2I$ revealed via weak photoexcitation

Yingchao Zhang[1*], Tika Kafle[1], Wenjing You[1], Xun Shi[1], Lujin Min[3], Huaiyu(Hugo) Wang[3], Na Li[1], Venkatraman Gopalan[3], Kai Rossnagel[7,8], Lexian Yang[5,6], Zhiqiang Mao[3,4], Rahul Nandkishore[2], Henry Kapteyn[1], Margaret Murnane[1]

[1]*Department of Physics and JILA, University of Colorado and NIST, Boulder, CO 80309, USA*
[2]*Department of Physics and Center for Theory of Quantum Matter, University of Colorado Boulder, Boulder, CO 80309, USA*
[3]*Materials Research Institute and Department of Materials Science & Engineering, The Pennsylvania State University, University Park, PA 16802, USA*
[4]*Department of Physics, The Pennsylvania State University, University Park, PA 16802, USA*
[5]*State Key Laboratory of Low Dimensional Quantum Physics, Department of Physics, Tsinghua University, Beijing 100084, China.*
[6]*Frontier Science Center for Quantum Information, Beijing 100084, China.*
[7]*Institute of Experimental and Applied Physics, Kiel University, D-24098 Kiel, Germany*
[8]*Ruprecht Haensel Laboratory, Deutsches Elektronen-Synchrotron DESY, D-22607 Hamburg, Germany*

\* Corresponding author: Yingchao.Zhang-1@colorado.edu

**This PDF file includes:**

Section 1. Error analysis of the observed flat band $(TaSe_4)_2I$ in Fig. 4
Section 2. Experimental and data analysis methods for time-resolved reflectivity results of $(TaSe_4)_2I$.
Fig S1. Energy distribution curves (EDC) of $(TaSe_4)_2I$ from 0.3 Å to 0.5 Å before pump at 80 K.
Fig S2. EDCs of the flat band of $(TaSe_4)_2I$ in Fig. 4 and corresponding error analysis
Fig S3. The dispersions of the bare bands at different time delays and the extraction of electron temperature at 200 fs via fitting Fermi-Dirac distribution
Fig S4. EDCs of $(TaSe_4)_2I$ at around $k_F$ before and after pump excitation at 80 K.
Fig S5. Time-resolved reflectivity results of $(TaSe_4)_2I$.
Fig S6. The dynamics of the in-gap spectral weight of $(TaSe_4)_2I$ within a region defined by the yellow dashed rectangle in Fig. 4C
Fig S7. trARPES mapping of $Rb_{0.3}MoO_3$
Fig S8. In-gap spectral weight dynamics of $(TaSe_4)_2I$ at different temperatures and pump fluences.

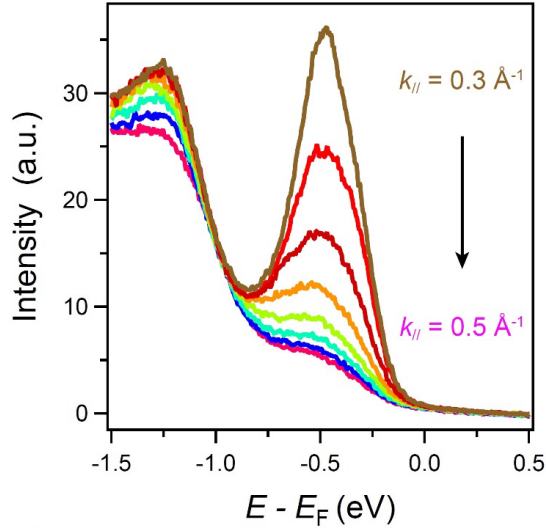

**Fig. S1.** Energy distribution curves (EDC) of $(TaSe_4)_2I$ from 0.3 Å to 0.5 Å before pump excitation at 80 K.

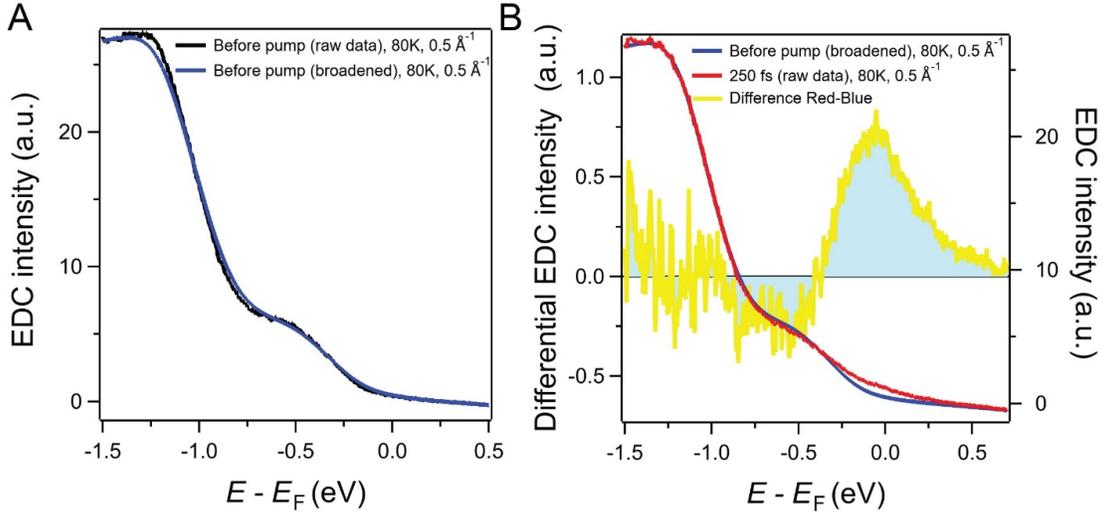

**Fig. S2.** (**A**) EDCs of $(TaSe_4)_2I$ before pump excitation at 80 K with (blue curve) and without Gaussian broadening (black curve). (**B**) **To right axis:** EDCs at 80 K before pump with Gaussian broadening (blue curve) and at 250 fs without Gaussian broadening (red curve). **To left axis**: the difference between red and blue curves (yellow curve).

## 1. Error analysis of the observed flat band $(TaSe_4)_2I$ in Fig. 4

At 0.5 Å which is away from $k_F$, new spectral weights appear around the Fermi level after pump excitation as shown in Fig. 4A-D. We claim in the main text that these new weights belong to the newly emerged flat band, consisting of numerous single-polaron subbands. Nevertheless, the broadening and shift of the broad hump at lower energy are also possible to present some extra spectra weights around the Fermi level. Here we provide further analysis to rule out this possibility by presenting the differential curve between EDCs at 250 fs and before pump. However, it's common for trARPES spectra to have slight inhomogeneous broadenings after pump excitations. The broadening mechanisms include Stark effect, increased carrier scatterings, pump-probe spatial cross-correlation and etc., which can be seen as extrinsic effects. If we directly subtract the black curve in (**A**) from the red curve in (**B**) to get the differential EDC, the extrinsic broadening will slightly contribute to the gain of spectral weights around Fermi level after pump excitation. Thus, in (**A**), we intentionally broaden the EDC before pump (black curve) by convolving it with a Gaussian function with 0.1 eV width, so that resulting blue curve has a good agreement with the raw EDC after pump (red curve) at the range from -1.5 eV to -0.8 eV. Then we take the difference between red and blue curves, as the yellow curve in (**B**). Note that the yellow curve shows a net gain of the spectral weight between -1 eV and 0.5 eV, which is a strong evidence that the broadening and shift of the broad hump at higher binding energy are not the major origins of the flat band around the Fermi level, as both of them conserve the spectral weight. The net gain in the differential EDC is persistent with and without the intentional broadening. The extrinsic broadening only slightly increases the intensity around the Fermi level, as shown in (**A**). The differential maps in Fig. 4 are from the direct subtractions of raw data without intentional broadening.

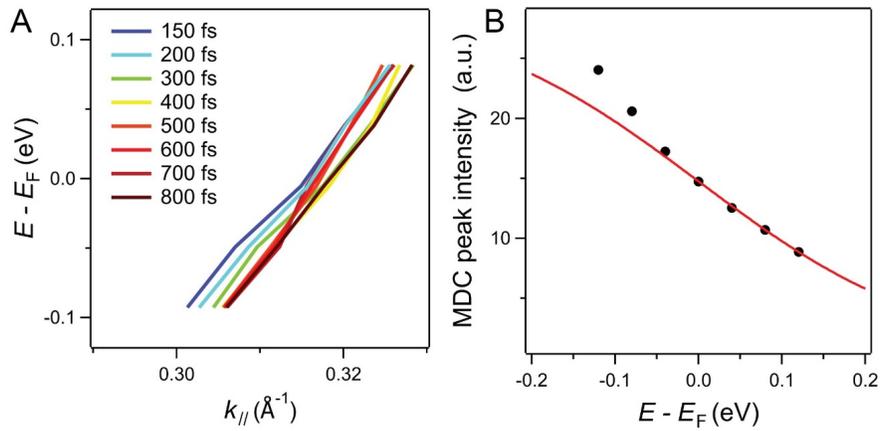

**Fig. S3. (A)** The dispersions of the bare bands of $(TaSe_4)_2I$ at different time delays at 300 K via fitting the peak positions of momentum distribution curves (MDC). Note that the band velocities at different time delays are basically the same. **(B)** The black dots are energy-dependent peak intensities of MDCs at 200 fs, while the red curve is Fermi-Dirac function at 1600 K temperature.

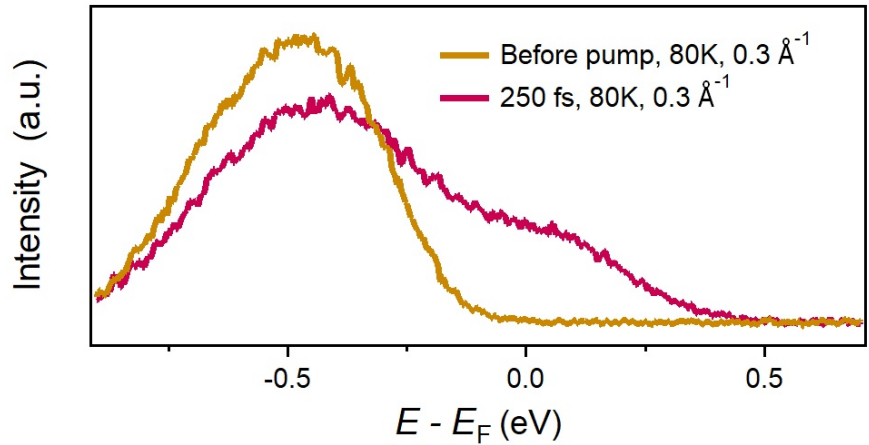

**Fig. S4. EDCs of $(TaSe_4)_2$ at around $k_F$ before and after pump at 80 K.** The second peak around Fermi level in the pink curve is the newly emerged bare band.

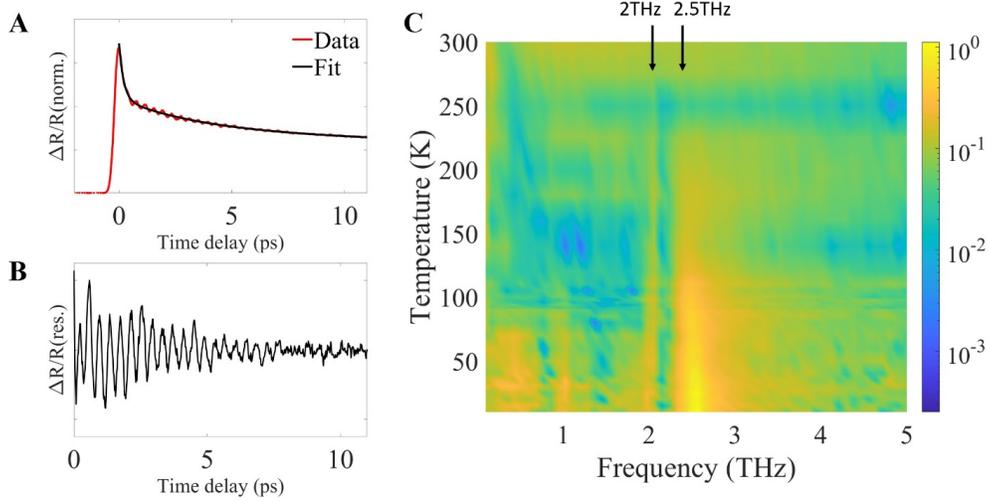

**Fig. S5. (A)** shows a typical fitting of the exponential decay baseline in optical pump optical probe data collected on $(TaSe_4)_2I$, the example shown is at 10K. **(B)** shows the residue between raw data and the fitting model, which represents the coherent phonon oscillations. **(C)** is the FFT color map of the extracted coherent oscillation as a function of temperature.

## 2. Experimental and data analysis methods for time-resolved reflectivity results of $(TaSe_4)_2I$.

The transient optical spectroscopy is collected with 800 nm ultrafast laser of 100 fs pulse width from MaiTai system with 80MHz repetition rate. The pump beam is generated from 400 nm SHG signal of a BBO crystal and the probe beam is 800 nm. The phenomenological model used for fitting the ΔR/R vs. delay time is the following:

$$\frac{\Delta R(t)}{R(t)} = I_{peak,1} * exp\left(\frac{t-t_0}{\tau_1}\right) + I_{peak,2} * exp\left(\frac{t-t_0}{\tau_2}\right) + I_{peak,3} * exp\left(\frac{t-t_0}{\tau_3}\right)$$

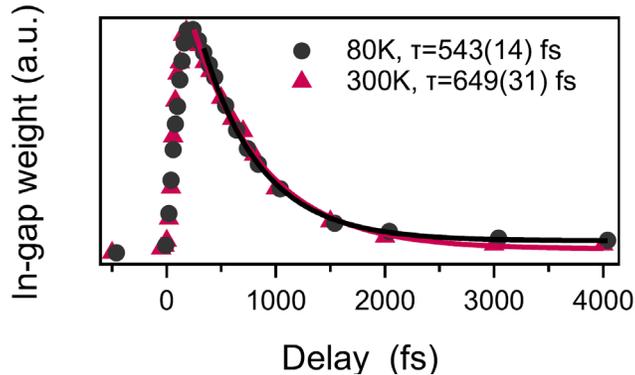

**Fig. S6.** The dynamics of the in-gap spectral weight of $(TaSe_4)_2I$ within a region defined by the yellow dashed rectangle in Fig. 4C at 80 K (black symbols) and 300 K (pink symbols). The blue dashed line determines the peak of in-gap spectral weight dynamics happening at ~250 fs. The solid lines are the exponential fitting curves from which the decay constant τ is extracted.

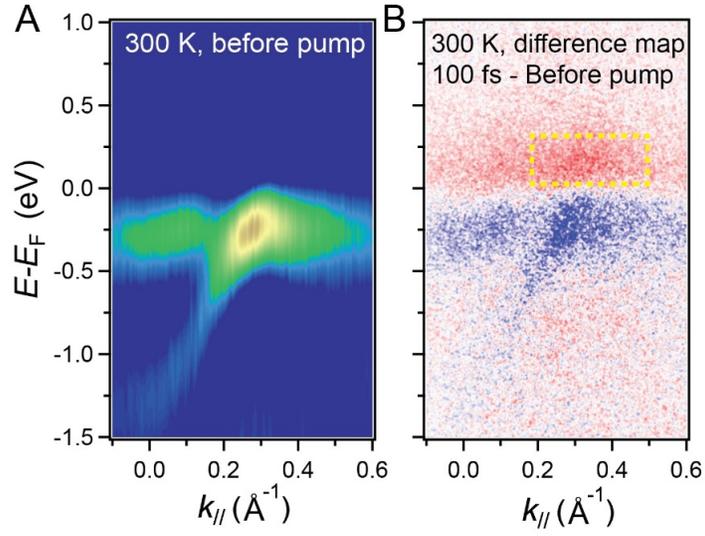

**Fig. S7.** (**A**) The second derivative image of the ARPES map of $Rb_{0.3}MoO_3$ along $\Gamma - Y$ direction at 300 K. (**B**) Difference map between 100 fs and before pump at 300 K with 0.6 mJ/cm$^2$ pump fluence. The yellow dashed rectangle denotes the region of interest mentioned in the main text and Fig. 4E.

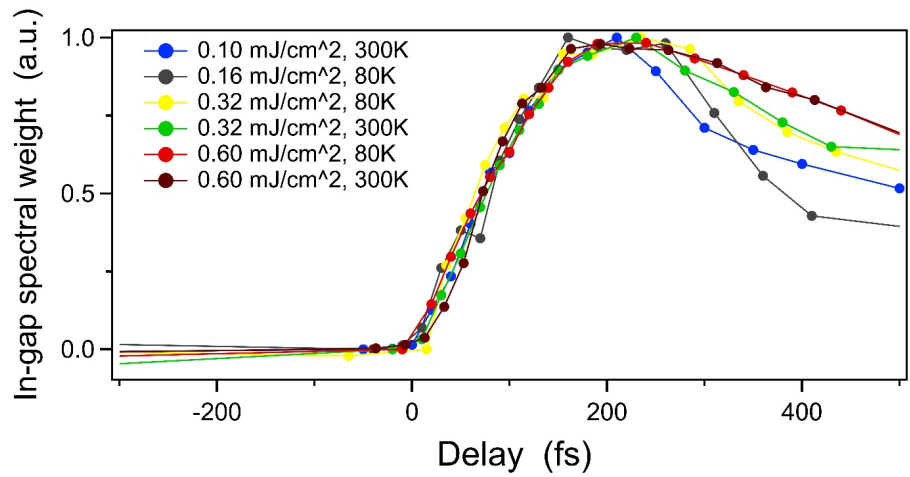

**Fig. S8.** In-gap spectral weight dynamics of $(TaSe_4)_2I$ at different temperatures and pump fluences.